\documentclass[a4paper,11pt]{article}

\usepackage[latin1]{inputenc}            
\usepackage{amsmath}                     % Matematiske kommandoer
\usepackage{amssymb}                     % Matematiske symboler
\usepackage{amsfonts}
\usepackage{latexsym}
\usepackage{graphicx}

\usepackage{dcolumn}
\usepackage{layout}
\usepackage{multirow}
\usepackage{longtable}
\usepackage{verbatim}
\usepackage{graphicx}
\usepackage[danish]{varioref}
\usepackage{eurosym}

\newtheorem{theorem}{Theorem}[section]

\newtheorem{remark}{Remark}[section]

\begin{document}
\Large
\begin{center}
Reserve-Dependent Surrender
\end{center}
\normalsize
\begin{center}
Kamille Sofie T\aa gholt Gad (1), Jeppe Juhl (2), Mogens Steffensen (1)\\
((1) University of Copenhagen, (2) Edlund A/S)
\end{center}

\begin{abstract}
We study the modelling and valuation of surrender and other behavioural
options in life insurance and pension. We place ourselves in between the two
extremes of completely arbitrary intervention and optimal intervention by
the policyholder. We present a method that is based on differential
equations and that can be used to approximate contract values when
policyholders exhibit optimal behaviour. This presentation includes
a specification of sufficient conditions for both consistency of the model
and convergence of the contract values. When not going to the limit in the
approximation we obtain a technique for balancing off arbitrary and optimal
behaviour in a simple, intuitive way. This leads to our suggestions for
intervention models where one single parameter reflects the extent of
rationality among policyholders. In a series of numerical examples we
illustrate the impact of the rationality parameter on the contract values.
\end{abstract}

\section{Introduction}
\label{intro}
Modern solvency and accounting rules (Solvency II and IFRS) require that
expected policyholder behaviour is taken into account. This includes e.g.
expected surrender and expected transcription into free policy (paid-up
policy). The expectation is supposed to take into account both the economic
conditions under which the behaviour takes place as well as the extent to
which intervention is to the benefit of the policyholder. The economic
conditions and how beneficial it is for the policyholder to intervene may
change over time. Therefore, one should properly speak of dynamic behaviour
models when formalizing these effects in the actuarial valuation formulas.
Changing economic conditions could e.g. be a changing level of interest
rates, and one idea would be to let the intensity or probability of
intervention depend on the current (possibly stochastic) level of interest rates. How
beneficial an action is can be formalized by the gain from intervention.
Determining the gain may be a delicate issue since both intervening and not
intervening opens up for new intervention options in the future that also
have to be taken into account. E.g., not to surrender typically opens up for
surrendering later, and transcription into free policy changes the effect of
the surrender option. This challenge calls for a recursive solution such
that the gain is always measuring correctly the tendencies of intervening in
the future. We disregard the economic condition by assuming deterministic
interest rates and focus on the latter idea of a recursive formula to deal
with the benefit of intervention. One motivation for this focus is that,
perhaps, the external economic conditions are supposed to approximate to the
internal benefit.

There exists a range of approaches to modelling of behavioural risk. One
extreme is to say that intervention occurs in a completely arbitrary way,
like insurance risk. We hereby mean that we model the behaviour as
independent of everything else in our model than the state of the
policyholder and the time measured through calendar time, the policyholder's
age, time since initiation, or time to (deterministic) retirement.
Specifically, the behaviour depends on neither the contract the policyholder
holds nor the interest rate. With this approach it is tractable to study
various aspects beyond just adding surrender to a survival model. Buchardt
et al. \cite{BMS13} studied the formalistic interaction between semi-Markov
modelling of insurance risk and behavioural risk, including duration
dependence of mortality and payments in the disability state and recognizing
duration dependence of free policy payments. A simpler exposition is found
in Buchardt and M\o ller \cite{BM13}. Henriksen et al. \cite{HNSS14} also combine
surrender and free policy options and study the impact on reserving from
different simplifying assumptions about the dependence between insurance
risk and behaviour risk.

Another extreme is to say that intervention occurs in a completely rational,
optimal way. We hereby mean that the policyholder, who is assumed to have the
same information as the insurance company has, intervenes according to a
strategy that maximizes the value of the insurance contract. This approach
was taken in Steffensen \cite{MS02}, who derived general variational inequalities
that characterize the reserve in case of a multi-state Markov model for
insurance risk and a multi-state model for behavioural risk. In the
surrender case, this is known as American option pricing of surrender risk.
Other early references based on this approach to surrender risk are Grosen
and J\o rgensen \cite{GJ00} and Bacinello \cite{Ba03}.

In between these extremes exist all different kinds of models where
intervention is modelled by an intensity, but where the intensity not only
depends on time but also some stochastic factors. The dependence on the
interest rate appears obvious and is thoroughly examined by De Giovanni
\cite{Gi10}, who calculate reserves by solving partial differential equations
numerically. There exists a large amount of literature examining relevant
explanatory variables but since these studies appear somewhat marginal to
our approach we refer to Eling and Kiesenbauer \cite{EK14} and references therein
for a comprehensive literature overview.

Rather than letting the intensity depend on external factors, one could let
the intensity depend on internal factors relevant to the specific policy.
That could e.g. be to take the difference between the surrender value and
(some notion of the) reserve as a measure of how beneficial an intervention
is. If the reserve compared with the surrender value does not take future
intervention options into account, the calculation can be split up in two
standard exercises: First, calculate the reserve without intervention and
then plug this reserve into the intensity for a calculation including
surrender. If the reserve compared with the surrender value does take future
intervention into account, the (usually) linear Thiele differential equation
characterizing the reserve becomes in general a non-linear differential
equation. The non-linear term comes from the risk premium with respect to
the surrender event that contains a non-linear function of the reserve
itself. The rationale for this paper is to take a thorough look at this
non-linear differential equation in order to motivate it, interpret it,
generalize it, and solve it numerically. Last but not least, we present a
probabilistic proof of and clarify sufficient conditions for a convergence
result that may seem intuitively clear: If the tendency to intervene tends
to zero whenever the gain from intervention is negative and tends to
infinity whenever the gain from intervention is positive, we reach in the
limit at the reserve based on completely rational behaviour. We establish
sufficient convergence of intensities to reach such a conclusion. Thus, our
approach to intervention option pricing has two purposes: First, it
represents in itself a relevant approach in between the two extremes that,
certainly, takes into account the extent to which intervention is to the
benefit of the policyholder. Second, for simple parametric forms of the
intensity, our calculation approximates the largest possible liability. As
such it can be used as a worst-case or stress calculation with respect to
surrender risk.

The idea of approximating the maximum value by a series of solutions to
differential equation has been known as the penalty method. In computational
finance it has been used as an approximation method for American option
pricing. In Forsyth and Vetzal \cite{FV02} the penalty method is compared with
alternative techniques for pricing of the American put option. In Gad and
Pedersen \cite{GP14} the modelling of non-rational
option holder behaviour is studied in a way similar to what is done here. The contribution of the present paper is three-fold: First, we introduce, to the
knowledge of the authors, for the first time the penalty method in
intervention option pricing in insurance. Second, we prove sufficient
conditions for the convergence to hold. Third, we do not only think of the
intensity model as a means of approximating the largest value, but as a
highly relevant approach to general intervention option pricing, useful in
accounting and solvency. The approach balances arbitrariness and benefit in
a simple form, and in some examples we catch the notion of rationality in
one single parameter.
\section{Standard Setup}
\label{sec:2}
\begin{figure}[h]
  \begin{picture}(10,30)
  \put(55,0){\framebox(100,20){Active}}
  \put(240,0){\framebox(100,20){Dead}}
  \put(155,10){\vector(1,0){85}}
  \put(190,15){$\mu(t)$}
  \end{picture}
\caption{Standard survival model.}
\label{fig:AD}
\end{figure}
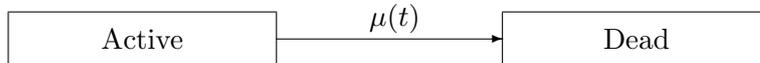

Consider a model with a policyholder who is either alive (active) or dead. We assume
the state of the policyholder is governed by a state process with a
deterministic, continuous death intensity, $\mu (t)$, see Figure \ref{fig:AD}%
. Let $I$ be the process indicating whether the policyholder is alive, and
let $N$ be the process counting the numbers of deaths of the policyholder.
The policyholder is assumed to have the following simple contract. She pays
a deterministic premium with continuous intensity $\pi (t)$ until a terminal
time, $n$, as long as she is alive. If she is alive at time $n$ she receives a
deterministic pension sum $b_a(n)$, and if she dies before then upon
death she gets a deterministic death sum, $b_{ad}(t)$. Thus, the accumulated
payments in the time interval $[0,t]$ is given by the following
process of accumulated payments: 
\begin{equation*}%\nonumber
B(t)=B(0)-\int_{0}^{t}\pi (u)I(u)du+\int_{0}^{t}b_{ad}(u)dN(u)+I(n)b_a(n)1_{(t\geq n)},
\end{equation*}
\noindent for $t\in \lbrack 0,n]$. We assume that the market offers a
deterministic, continuous interest rate, $r(t)$. We introduce the reserve
corresponding to the policyholder being active as the conditional expected
present value of future payments,
\begin{equation}\nonumber
V(t)={\mathbb{E}}\left[ \left. \int_{t}^{n}e^{-\int_{t}^{u}r(\tau )d\tau
}dB(u)\right\vert I(t)=1\right] .
\end{equation}
We then know, e.g. from \cite{TMMS07}, that the reserve, $V$, is
continuously differentiable on $[0,n)$ and that it is the solution to
Thiele's differential equation,
\begin{equation}
V^{\prime }(t)=r(t)V(t)+\pi (t)-\mu (t)(b^{ad}(t)-V(t)),  \label{eq:thiele}
\end{equation}
with $V(n-)= b_a(n)$.

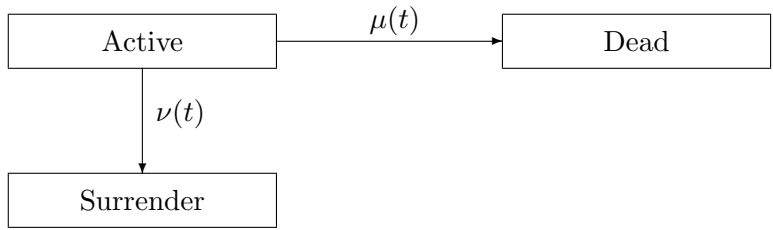
\begin{figure}[h]
	\begin{picture}(10,90)
	\put(55,0){\framebox(100,20){Surrender}}
	\put(105,60){\vector(0,-1){40}}
	\put(110,40){$\nu(t)$}
	\put(55,60){\framebox(100,20){Active}}
	\put(240,60){\framebox(100,20){Dead}}
	\put(155,70){\vector(1,0){85}}
	\put(190,75){$\mu(t)$}
	\end{picture}
\caption{Standard surrender model.}
\label{fig:ADS1}       
\end{figure}

We now add to our model the possibility that the policyholder surrenders.
That is, we add the possibility that the policyholder terminates
her contract and instead receives a deterministic, continuous surrender
value, $G(t)$. This can, e.g., be added to the model by assuming that the
policyholder at any time surrenders with some deterministic, continuous
intensity, $\nu (t)$, see Figure \ref{fig:ADS1}. We use the term active for when the policy is in force.

Mathematically, the state of surrender is in this model not different from
the state of death, except that the associated payments are different. The
reserve, $V_{\nu }$, is continuously differentiable and solves the following
Thiele's differential equation, see e.g. \cite{TMMS07},
\begin{equation}
V_{\nu }^{\prime }(t)=r(t)V_{\nu }(t)+\pi (t)-\mu (t)\left( b^{ad}(t)-V_{\nu
}(t)\right) -\nu (t)\left( G(t)-V_{\nu }(t)\right) ,  \label{eq:thieleg1}
\end{equation}%
with $V_{\nu }(n-)= b_a(n)$. 

For $V_{v}$ to be continuously differentiable we need, in general, that $\nu 
$ is continuous as assumed above. However, what is really needed is that $%
\nu (t)\left( G(t)-V_{\nu }(t)\right) $ is continuous and this can be
obtained even when $\nu $ is discontinuous and properly defined at the point
where $G(t)=V_{\nu }(t)$.

The surrender value $G$ can be anything exogenously given. In practice it
is, typically, a technical value of the same payment stream based on
technical assumptions on interest rates and intensities that we denote by $%
\left( r^{\ast },\mu ^{\ast }\right) $. In that case, the surrender value is
the technical reserve $V^{\ast }$ that solves (\ref{eq:thiele}) with $\left(
r,\mu \right) $ replaced by $\left( r^{\ast },\mu ^{\ast }\right) $.

\section{Reserve Dependent Surrender}
\label{sec:rds}
The forthcoming Solvency II regulations requires that the traditional
modelling of surrender is revisited. In Article 79 of the Solvency II Directive it is stated that 
\textit{"Any assumptions made by insurance and reinsurance undertakings with
respect to the likelihood that policyholders will exercise contractual
options, including lapses and surrender, shall be realistic and based on
current and credible information. The assumptions shall take account, either
explicitly or implicitly, of the impact that future changes in financial and
non-financial conditions may have on the exercise of those options"}. Thus,
we need to investigate and model what influences the policyholders choice to
surrender and we need to be able to calculate the reserves in the more
advanced models. In the present section we suggest a way to do this, and
discuss our method.

In a more realistic model of surrender we want to be able to express both
that surrender is likely influenced by how profitable it is, but also that
it is still random. On one hand, we wish surrender to be influenced by how
profitable it is, because surrender is a decision the policyholder makes. On
the other hand we also have multiple reasons for surrender being random.
Randomness is natural because the policyholder most likely lacks information
to decide what is profitable. Even if she had all the information that
the pension fund has and were able to use it, then her preferences may
differ seemingly randomly from the model set up by the pension fund because
of the policyholders personal preference and economical situation. She might
shift her job and get an offer from a new pension fund or she might
need cash for a divorce.

We can obtain randomness in our model by keeping the surrender modelled
by an intensity. Further, we model that the policyholders decision depends on
how profitable it is by letting the surrender intensity depend on how
profitable it is for the policyholder to surrender. If she surrenders at time 
$t$ she gains $G(t)$, but she loses the rest of the contract including her
right to exercise later. Hence, she loses $V_{\nu }(t)$. Therefore, we denote
by $G(t)-V_{\nu }(t)$ her profit from surrendering at time $t$. We would
like the surrender intensity to be non-negative and increasing in this
profit. At first glance this modelling seems to have a problem that the definition of the
surrender intensity is circular. However, Theorem \ref{thm:surDE} below
gives sufficient conditions for this circular definition not to be a
problem.
\begin{theorem}
\label{thm:surDE} For some given non-negative function, $h$, consider the
following differential equation in the function $U$: 
\begin{equation}  \label{eq:thieleint1}
U^{\prime}(t)= r(t)U(t)+\pi(t)-\mu(t)(b_{ad}(t)-U(t))-h(t,U(t))(G(t)-U(t)),
\end{equation}
with $U(n-)=\Delta B(n)$. Suppose (\ref{eq:thieleint1}) has a unique
solution, $U$, and define a surrender intensity by $\nu(t)\equiv h(t,U(t))$.
Then $U$ is the reserve when the policyholder chooses to surrender at time $%
t $ with intensity $\nu(t)$.
\end{theorem}
Proof: The possible problem in this model is the circular definition of the
surrender intensity. However, the existence and uniqueness of the solution
to both (\ref{eq:thieleg1}) and (\ref{eq:thieleint1}) ensures that this does
not become a problem. 

The process $\nu $ defined by $\nu (t)\equiv h(t,U(t))$
is uniquely determined from (\ref{eq:thieleint1}) and the reserve is then
uniquely determined from (\ref{eq:thieleg1}). It follows from the definition
of $U$ that $U$ solves (\ref{eq:thieleg1}), and then from the uniqueness of
the solution to (\ref{eq:thieleg1}) it follows that the reserve is given by $%
U$.

Once we have decided on a policyholder with a specific policy and a function 
$h$, and thereby also $\nu $, then for this single policyholder, our model
does not differ from a model with a deterministic time dependent surrender
intensity as what we had in the classical model of (\ref{eq:thieleg1}).
However, when we use the model for pricing a portfolio of insurance
contracts for a group of policyholders, then the model assigns different
surrender intensities to each policyholder. Thereby, the reserves in general
become higher than if we had used a constant surrender intensity or a
specific time dependent surrender intensity for the whole portfolio. 

The relation between the surrender intensity and the profitability may be
chosen in many different ways. Two examples we investigate are: 
\begin{equation}
\nu (t)=h(t,V_{\nu }(t))=\psi \exp \{\theta (G(t)-V_{\nu }(t))\},
\label{eq:surintexp}
\end{equation}
\begin{equation}
\nu (t)=h(t,V_{\nu }(t))=\theta 1_{(G(t)-V_{\nu }(t)>0)},
\label{eq:surintabs}
\end{equation}
where $\psi $, $\theta >0$ are constants. For equation (\ref{eq:surintexp}), 
$\psi $ tells about the overall tendency to surrender, whereas $\theta $
tells about how profitability creates deviations from this tendency. For
equation (\ref{eq:surintabs}), $\theta $ controls both. In both cases we
speak of $\theta $ as the rationality parameter. Other intensity functions
can be chosen and one should choose a functional form which matches with
data. The only mathematical requirement is that the function $h$ has to make
it possible to use Theorem \ref{thm:surDE}. 

One immediate drawback of our model is that we most often do not have an
explicit solution for the differential equation (\ref{eq:thieleint1}). This
implies that we do not have an explicit expression for the reserve. However,
we do have algorithms available for numerical solutions to ordinary
differential equations.

\section{Reserve Dependent Policyholder Behaviour}
\label{sec:4}
\begin{figure}[h]
	\begin{picture}(10,90)
	\put(150,0){\framebox(100,20){Free Policy}}
	\put(200,60){\vector(0,-1){40}}
	\put(150,70){\vector(-2,-1){40}}
	\put(150,10){\vector(-2,1){40}}
	\put(250,70){\vector(2,-1){40}}
	\put(250,10){\vector(2,1){40}}
	\put(205,40){$\nu_{af}(t)$}
	\put(100,70){$\nu_{as}(t)$}
	\put(260,70){$\mu(t)$}
	\put(100,10){$\nu_{fs}(t)$}
	\put(260,10){$\mu(t)$}
	\put(150,60){\framebox(100,20){Active}}
	\put(50,30){\framebox(100,20){Surrender}}
	\put(250,30){\framebox(100,20){Dead}}
	\end{picture}
\caption{Free policy and surrender model.}
\label{fig:ADSF}    
\end{figure}
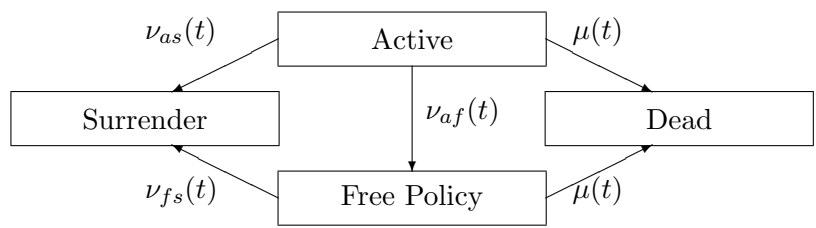

The idea of modelling behaviour by profit dependent intensities may be used
for other applications as well. Within life insurance the policyholder's
choice to convert into free policy (paid-up policy) has some resemblance
with the surrender choice. Thereby we may find it reasonable to expand our
model with the possibility of conversion into free policy in the same way as
we added surrender. Figure \ref{fig:ADSF} displays a simple model where $\nu
_{af}$ denotes the intensity of conversion into free policy, $\nu _{as}$
denotes the intensity of surrender when active and $\nu
_{fs}$ denotes the intensity of surrender after converted into free policy. Here the term active is used when the policyholder is paying premiums.

If all transition intensities are known explicitly, this model is studied in \cite{HNSS14}. When a policyholder converts into free policy the payments are reduced depending of the time of conversion. Let $b_{fd}(t,u)$ denote the death sum at time $t$ if
converted into free policy at time $u$, let $b_f(n,u)$ denote the terminal
payment at time $n$ if converted into free policy at time $u$, and let $G_f(t,u)$ denote the surrender value at time $t$ when converted into free policy at time $u$.
For the reserves we let $V_a(t)$ denote the reserve at time $t$ if the policyholder is active, and let $V_f(t,u)$ denote the reserve at time $t$ if the policyholder is in the free policy state and converted to free policy at time $u$.
 Now, we assume that the intensities are reserve
dependent and given in the form 
\begin{eqnarray*}
\nu _{as}(t) &=&h_{as}(t,V_{a}(t)), \\
\nu _{af}(t) &=&h_{af}(t,V_{a}(t)), \\
\nu _{fs}(t,u) &=&h_{fs}(t,u,V_{f}(t,u)).
\end{eqnarray*}
Then the reserves are given from the following differential equations: 
\begin{eqnarray*}
\frac{\mathrm{d}}{\mathrm{d} t}V_{a}(t) &=&r(t)V_{a}(t)+\pi (t)-\mu
(t)(b_{ad}(t)-V_{a}(t)) \\
&&-h_{as}(t,V_{a}(t))(G(t)-V_{a}(t))-h_{af}(t,V_{a}(t))(V_{f}(t,t)-V_{a}(t)),
\\
V_{a}(n-) &=&b_a(n),\\
\frac{\partial }{\partial t}V_{f}(t,u) &=&r(t)V_{f}(t,u)-\mu
(t)(b_{fd}^*(t,u)-V_{f}(t,u)) \\
&&-h_{fs}(t,u,V_{f}(t,u))(G^f(t,u)-V_{f}(t,u)), \\
V_{f}(n-,u) &=&b_f(n,u).
\end{eqnarray*}
The only requirement is that the system of differential equations has a unique
solution. However, the differential equations from above are heavy to work with, as we need to solve a new differential equation for each value of $V_f(t,t)$. When modelling the free policy option, this problem is usually overcome by introducing a scaling function, $f$, that describes the reduction of payments as a result of the conversion to free policy. Thus, $b_{fd}(t,u)=f(u)b_{ad}(t)$, $b_f(n,u)=f(u)b_a(n)$ and $G_f(t,u)=f(u)G_f(t)$. Assume the transition intensity $\nu_{fs}$ does not depend on the time of transition to free policy. Then the prospective reserve, $V_f^*(t)$, from the free policy state based on the payments $G_f(t)$, $b_{ad}(t)$ and $b_a(n)$ does not depend on this transition time either, and we get $V_f(t,u)=f(u)V_f^*(t)$ with
\begin{eqnarray*}
\frac{\mathrm{d}}{\mathrm{d} t}V_{f}^*(t) &=&r(t)V_{f}^*(t)-\mu
(t)(b_{ad}(t)-V_{f}^*(t)) \\
&&-\nu_{fs}(t)(G^f(t)-V_{f}^*(t)), \\
V_{f}^*(n-) &=&b_a(n).
\end{eqnarray*}
This makes $V_f(t,u)$ a lot easier to calculate. For more on the determination of the reference payments and scaling function, see \cite{HNSS14}. Note however that if $\nu_{fs}$ cannot depend on the time of transition to free policy, $u$, then it cannot depend on $G_f(t,u)-V_f(t,u)$ either and this is a large disadvantage.

To get profit dependent choices we may use 
\begin{eqnarray*}
\nu _{as}(t) &=&h_{as}(t,V_{a}(t))=\psi _{as}e^{\theta
_{as}(G(t)-V_{a}(t))}, \\
\nu _{af}(t) &=&h_{af}(t,V_{a}(t))=\psi _{af}e^{\theta
_{ap}(V_{f}(t,t)-V_{a}(t))}, \\
\nu _{fs}(t,u) &=&h_{fs}(t,u,V_{f}(t,u))=\psi _{fs}e^{\theta
_{fs}(G_f(t,u)-V_{f}(t,u))}.
\end{eqnarray*}

\section{Approximation of the Worst Case Reserve}
\label{sec:5}
In the two previous sections we discussed our model with reserve dependent surrender and we found it being a reasonable model for predicting the dynamics of surrender. However, in the following section we discuss how the model may also be used for determining worst case reserves when the true dynamics of the surrender intensity is not known. This is because our model is a version of what in the literature is known as the penalty method, and a large rationality parameter gives us the worst case reserve.

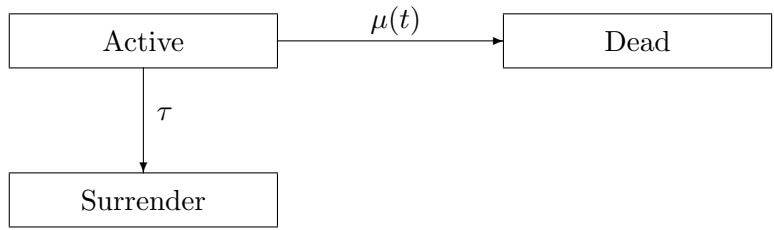
\begin{figure}[h]
	\begin{picture}(10,90)
	\put(55,0){\framebox(100,20){Surrender}}
	\put(105,60){\vector(0,-1){40}}
	\put(110,40){$\tau$}
	\put(55,60){\framebox(100,20){Active}}
	\put(240,60){\framebox(100,20){Dead}}
	\put(155,70){\vector(1,0){85}}
	\put(190,75){$\mu(t)$}
	\end{picture}
\caption{Optimal surrender model.}
\label{fig:ADS2}     
\end{figure}

Typically the technical reserve is paid out upon surrender (potentially minus expenses). In that case, if we take maximum of the technical reserve and the market reserve calculated under the assumption of no surrender, then we get a worst case reserve of either surrendering immediately or never surrender. However, a surrender strategy somewhere in between the two extremes may result in a higher market reserve. For determining the worst case reserve we consider all possible surrender strategies. To do this we construct a more general model. We assume that the transition from
active to surrender is governed by a randomized stopping time, $\tau $, with
respect to the state of the policyholder, with randomized stopping times
being defined as in \cite{ANS78}. That is, the time of surrender may depend on everything but the future time of death and the future interest rate. If the policyholder never surrenders her
contract we let $\tau =n$. The model is illustrated in Figure \ref{fig:ADS2}. The class of admissible surrender strategies at time $t$ are the variables in $[t,n]$ that are randomized stopping times with respect to the
filtration generated from $I$. We denote this class by $\mathcal{T}_{t}$.

We hereby disregard the possibility that the policyholder has more information about her future time of death than the insurance company has. We do this despite that such knowledge could influence the policyholders decisions.

Let $V_{\tau }$ denote the prospective reserve if the policyholder surrenders according to the randomized stopping time $\tau$. Assume $G(n)=0$, assume $G(n-)\leq V(n-)$ and assume $G$ continuous on $[0,n)$. Then from \cite{TMMS07} it follows that $V_{\tau }$ is given by: 
\begin{eqnarray*}
V_{\tau }(t) &=&{\mathbb{E}}\left[ \left. \int_{t}^{\tau
}e^{-\int_{t}^{u}r(x)dx}dB(u)+e^{-\int_{t}^{\tau }r(x)dx}G(\tau )I(\tau
)\right\vert I(t)=1\right]  \\
&=&V(t)+{\mathbb{E}}\left[ \left. e^{-\int_{t}^{\tau }r(x)dx}G(\tau )I(\tau
)-\int_{\tau }^{n}e^{-\int_{t}^{u}r(x)dx}dB(u)\right\vert I(t)=1\right]  \\
&=&V(t)+{\mathbb{E}}\left[ \left. e^{-\int_{t}^{\tau }r(x)dx}I(\tau )
G(\tau ) \right\vert I(t)=1 \right]  \\
&&-{\mathbb{E}}\left[ \left. e^{-\int_{t}^{\tau }r(x)dx}I(\tau )
{\mathbb{E}}\left[ \left. \int_{\tau }^{n}e^{-\int_{\tau
}^{u}r(x)dx}dB(u)\right\vert \tau, I(\tau )=1\right]  \right\vert I(t)=1 \right]  \\
&=&V(t)+{\mathbb{E}}\left[ \left. e^{-\int_{t}^{\tau }r(x)dx}I(\tau )\left(
G(\tau )-V(\tau )\right) \right\vert I(t)=1\right] .
\end{eqnarray*}
Consider the worst case scenario for the pension fund, where the policy
holder chooses the surrender strategy as the stopping time, $\tau$, that
maximizes $V_{\tau }$. This is an optimal stopping problem. Any classical
stopping time from the filtration generated by $I$ must fulfil $\tau I(\tau) =t_{0}I(t_{0})$ for some deterministic $t_{0}\in
[ t,\infty]$. The reserve is then given by: 
\begin{eqnarray*}
V_{\tau }(t) &=&V(t)+{\mathbb{E}}\left[ \left. e^{-\int_{t}^{\tau
}r(x)dx}I(\tau )\left( G(\tau )-V(\tau )\right) \right\vert I(t)=1\right]  \\
&=&V(t)+e^{-\int_{t}^{t_{0}}r(x)+\mu (x)dx}\left( G(t_{0})-V(t_{0})\right) .
\end{eqnarray*}
Thus, for the classical optimal stopping problem, without randomization
allowed, it is optimal to choose $t_{0}$ as any time from the set: 
\begin{equation}\nonumber
A_{t}\equiv \arg \max_{u\in \lbrack t,n]}\left( e^{-\int_{t}^{u}r(x)+\mu
(x)dx}(G(u)-V(u))\right) .
\end{equation}
As the inner part is continuous in $u$ on $[0,n)$ and as $G(n-)-V(n-)<G(n)-V(n)$, then $A_{t}$ must have a largest element.
Denote this element by $u^{\ast }$, i.e. let $u^{\ast }(t)\equiv \max A_{t}$%
, such that $u^{\ast }(t)$ is the latest optimal time to surrender. Let $%
\tau ^{\ast }=u^{\ast }(t)I(u^{\ast }(t))+n(1-I(u^{\ast }(t)))$. We define
the worst case reserve, $W$, by: 
\begin{equation}\nonumber
W(t)=\sup_{\tau \in \mathcal{T}_{t}}V_{\tau }(t)=V_{\tau ^{\ast }}(t).
\end{equation}
By a proof similar to the one of the verification theorem of Chapter 9 of 
\cite{KYPR06} it may be seen that $\tau ^{\ast }$ is optimal even if we
allow randomized surrender strategies.

Now, assume a family of functions, $h_\theta$, is implied. Let $\tau_\theta$ denote the surrender strategy of surrendering at time $u$ with intensity $\nu_\theta(u)=h_\theta(G(u)-V_{\nu_\theta}(u))$ and let $V_\theta=V_{\nu_\theta}$ with $V_{\nu_\theta}$ as defined in Sect. ~\ref{sec:rds}. Let 
\begin{equation}\nonumber
\bar{h}_\theta(x)\equiv \sup_{y\leq x}h_\theta(y),
\end{equation}
and 
\begin{equation}\nonumber
\underline{h}_\theta(x)\equiv \inf_{y\geq x}h_\theta(y).
\end{equation}
Now, the following holds:
\begin{theorem}
\label{thm:conv} Suppose that for each $\theta \geq 0$ we have that $h
_{\theta }$ is defined in a way such that we may use Theorem \ref{thm:surDE}
and suppose that the surrender value $G$ is continuous on $[0,n)$ with $G(n-)\leq V(n-)$ and $G(n)=0$. Also assume for $x<0$%
: 
\begin{equation}
\bar{h}_{\theta }(x)\rightarrow 0,\ \ \ \theta \rightarrow \infty ,
\label{eq:conv1}
\end{equation}
and for $x>0$: 
\begin{equation}
\underline{h }_{\theta }(x)\rightarrow \infty ,\ \ \ \theta \rightarrow
\infty .  \label{eq:conv2}
\end{equation}
Then, for every $t\in \lbrack 0,n]$: 
\begin{equation}\nonumber
V_{\theta }(t)\rightarrow W(t),\ \ \ \ \ \theta \rightarrow \infty .
\end{equation}
\end{theorem}
For a proof, see the appendix.

\begin{remark}
Some of the details in the proof has been omitted, but a fully detailed proof following the same reasoning for a closely related result for an American Put option may be found in \cite{GP14}. The fact that the penalty method provides convergence and the rate of convergence is not new. However, we find the proof of our article and of \cite{GP14} interesting. This is because they visualize how the error terms may be thought of as probabilities of economically bad choices of the policyholder times the loss the policyholder faces from her bad choices.
\end{remark}

\section{Numerical Examples}
\label{sec:6}
In this section we show four examples of how various surrender models impact
the development of the reserves in four different interest rate situations.
In each example we consider a contract with a constant premium intensity of $%
\pi=7,000$, a death sum of $b^{ad}=1,000,000$  and a pension sum of $%
2,000,000 $ . All values measured in DKK. These numbers are chosen as they have a realistic level for a Danish pension policy. For fairly realistic numbers in EUR divide by ten. The policyholder is assumed to be 35 years old at time 0 and
the time of retirement is at age 65. Time is measured in years and her death
intensity is assumed to be given by: 
\begin{equation}\nonumber
\mu(t)=0.0005+10^{5.728-10+0.038*(t+35)}.
\end{equation}
This is the death intensity from the Danish life table G82 for females.
If the policyholder surrenders her contract, she receives a surrender value
given by the technical reserve. The technical reserve is based on the same
payments as the contract, and on a technical interest rate intensity of $%
\hat{r}=0.05$. Interest rates are chosen high to better visualize the impact of the choice of surrender intensity. We assume no extra expenses at surrender. Thus, the surrender
value is given from the differential equation: 
\begin{equation}
G^{\prime}(t)= \hat{r}G(t)+\pi-\mu(t)(b^{ad}-G(t)),
\end{equation}
with $G(n-)=2,000,000$. We consider the following five surrender models:
 
\begin{equation}
\begin{array}{llcl}
\textrm{Model a}: & \nu^a(t) & = & 0.05
\cdot\exp\{0.000003(G(t)-V_{\theta,\psi}(t))\} \nonumber\\ 
\textrm{Model b}: & \nu^b(t) & = & 0.05\cdot 1_{(G(t)-V_{\theta}(t))} \nonumber\\ 
\textrm{Model c}: & \nu^c(t) & = & 0.05 \nonumber\\ 
\textrm{Model d}: & \nu^d(t) & = & 0 \nonumber\\ 
\textrm{Model e}: & \nu^e(t) & = & 5\cdot1_{(G(t)-V_{\theta}(t))}.\nonumber
\end{array}
\end{equation}
The first three models are based on a surrender intensity of around 5\%. The last model is a model with a rationality parameter $\theta=5$, which has been found to be high enough for us to approximate the worst case reserve. 
Additionally, we consider four different developments of the interest rate, $r$, used
for pricing market reserves. For the two first interest rate situations we
compare the surrender value and the reserves for the five different
surrender models. For the two last interest rate situations we compare the
surrender value and the reserves for surrender Model d and Model e.
\begin{figure}[h]
  \includegraphics[width=0.95\textwidth]{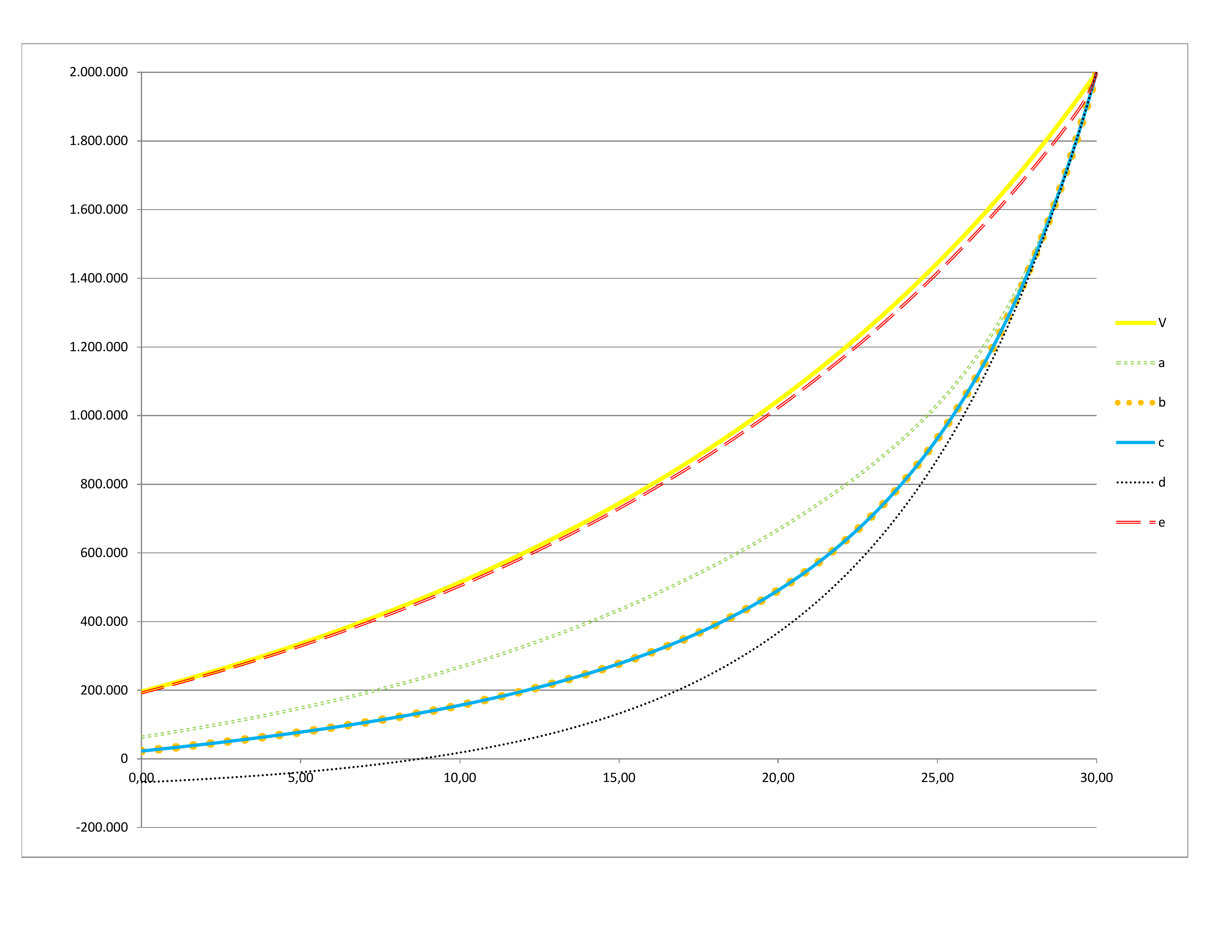}
\caption{Example 1. The technical interest rate is $\hat{r}=0.05$. The
market interest rate is $r=0.12$. Immediate surrender is always optimal.}
\label{fig:model1}      
\end{figure}
\paragraph{Example 1: Market interest rate is above technical interest
rate}
Assume $r=0.12$. The reserve developments are displayed in Figure \ref%
{fig:model1}. In this situation it is at all time points optimal for the
policyholder to surrender. The worst case reserve corresponds to the
surrender value. The lowest reserve is the market reserve based on no
surrender, Model d. Models with a chance of surrender has reserves in
between. Since there is no risk of surrendering too early, then Model b and
the traditional Model c do not differ. For Model a we get a slightly higher
reserve than the one for Model b and Model c, because the basic intensity $
0.05$ is slightly increased at all time points by the exponential factor in
the intensity.

\begin{figure}[h]
  \includegraphics[width=0.95\textwidth]{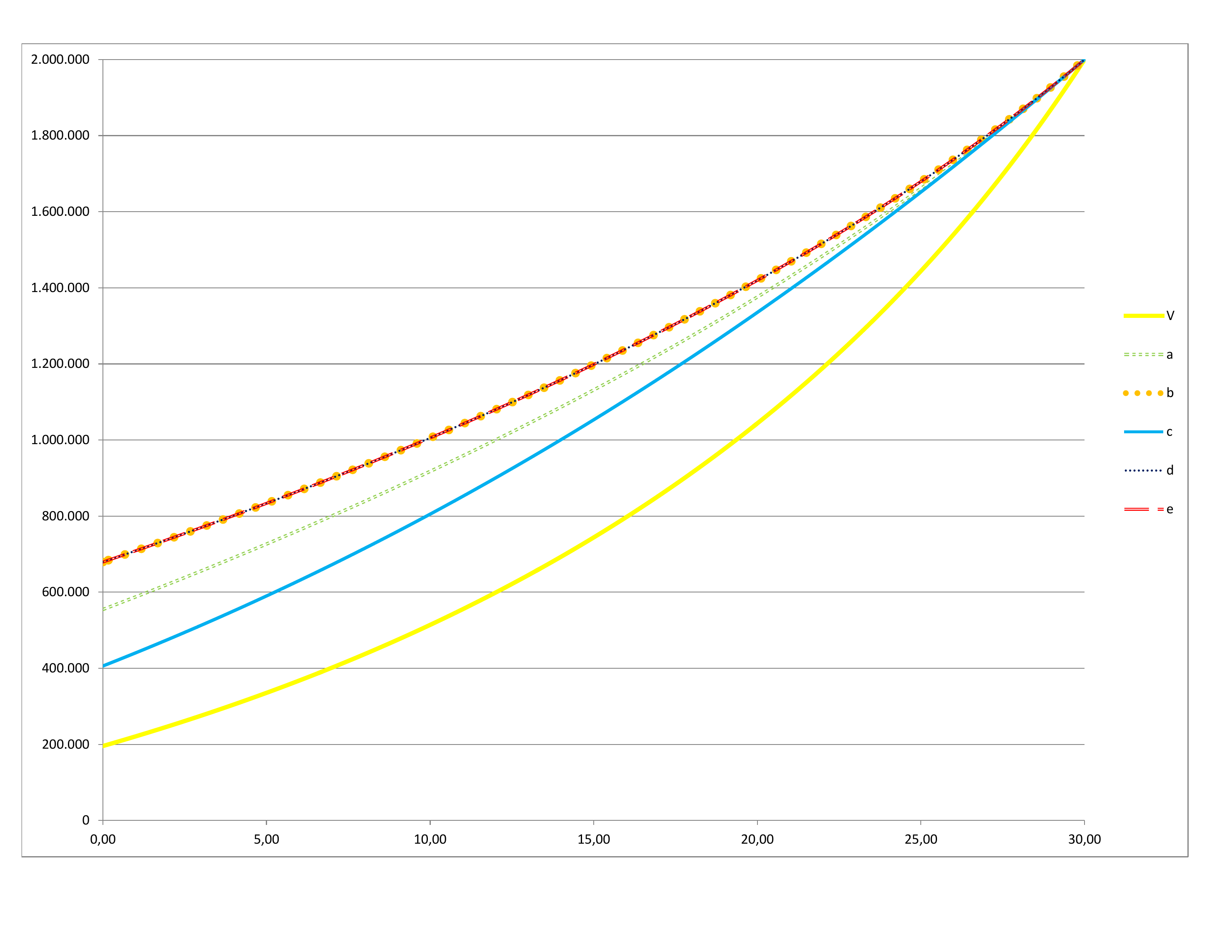}
\caption{Example 2. The technical interest rate is $\hat{r}=0.05$. The
market interest rate is $r=0.02$. Surrender is never optimal.}
\label{fig:model2}       
\end{figure}

\paragraph{Example 2: Market interest rate is below technical interest
rate}
Assume $r=0.02$. The reserve developments are displayed in Figure \ref
{fig:model2}. In this situation it is never optimal for the policyholder to
surrender. The worst case reserve corresponds to the market reserve with no
surrender. In Model b and Model e the policyholder does not make the mistake
of surrendering if it is not profitable, and thus, this has an equally high
reserve. The surrender value is the lowest value and the reserves of Model a
and the traditional Model c are in between. Model a has a higher reserve
than Model c, because the basic intensity $0.05$ is slightly increased at
all time points by the exponential factor in the intensity. 

\begin{figure}[h]
  \includegraphics[width=0.95\textwidth]{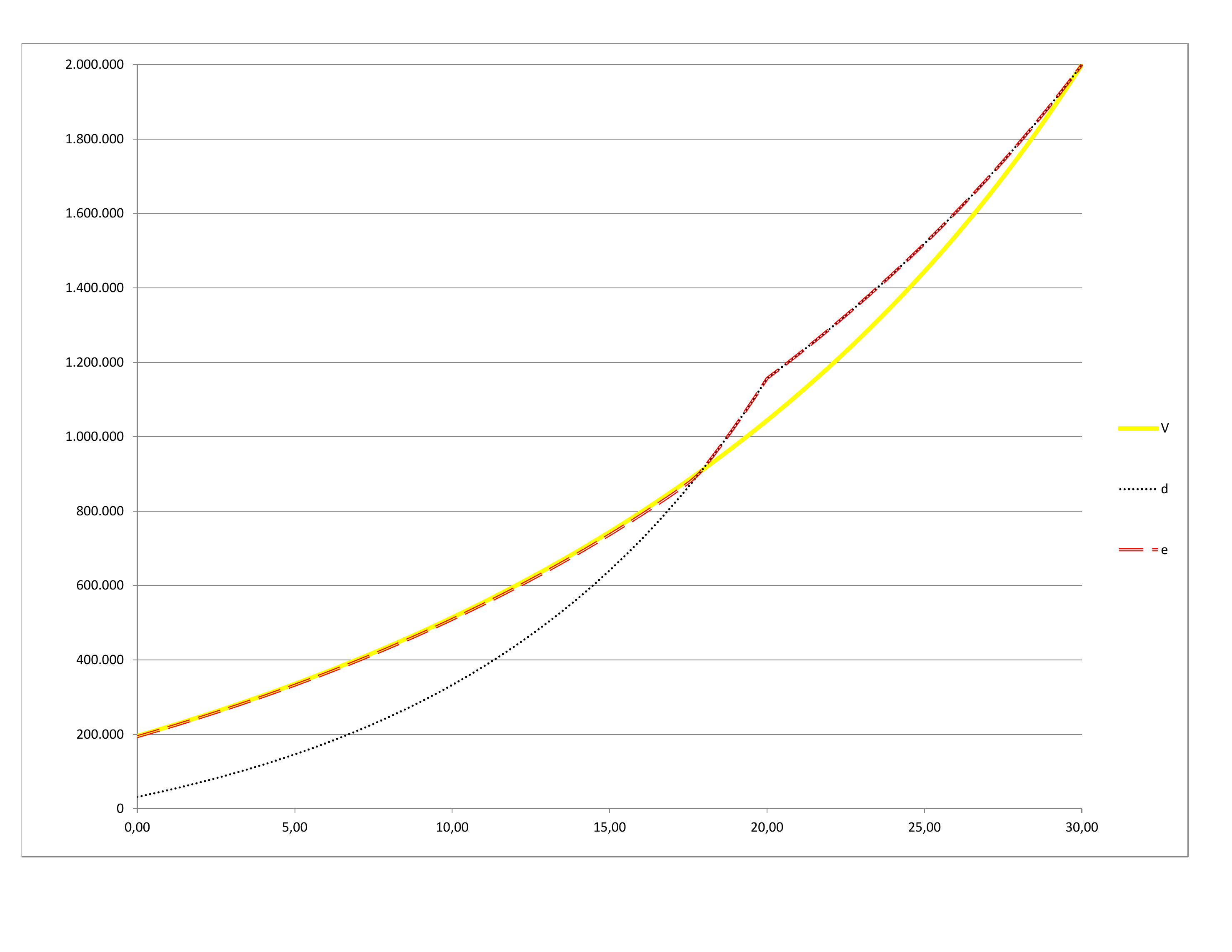}
\caption{Example 3. The technical interest rate is $\hat{r}=0.05$. The
market interest rate is $r(t)=0.10\cdot 1_{(t\leq 20)}+0.04\cdot 1_{(t>20)}$. Surrender is optimal if the surrender value is higher than the market reserve
with no surrender.}
\label{fig:model3}   
\end{figure}

\paragraph{Example 3: Market interest rate is decreasing} 
Assume $r(t)=0.10\cdot 1_{(t\leq 20)}+0.04\cdot 1_{(t>20)}$. The qualitative feature we capture is that the interest rate crosses the guaranteed interest rate downwards. The reserve
developments are displayed in Figure \ref{fig:model3}. In this situation it
is optimal to surrender if the surrender value is higher that the market
reserve in Model d with no surrender. Thus, after time $t=20$ it is optimal
to keep the contract because the technical interest rate is higher than the
market interest rate. Right before time $t=20$ the interest rate of the
market is higher than the technical interest rate, but this\ is only for a
short time, and thus it is still optimal to keep the policy in order to
benefit from the technical interest rate later on. At some point before time 
$t=20$ the surrender value and the market reserve of Model d intersects.
Before this time it is optimal to surrender because the gain from the high
market interest rate before time $t=20$ is then higher than the future loss
from the low market interest rate. All together the worst case reserve is
given as the maximum of the surrender value and the market reserve with no
surrender. 

\begin{figure}[h]
  \includegraphics[width=0.95\textwidth]{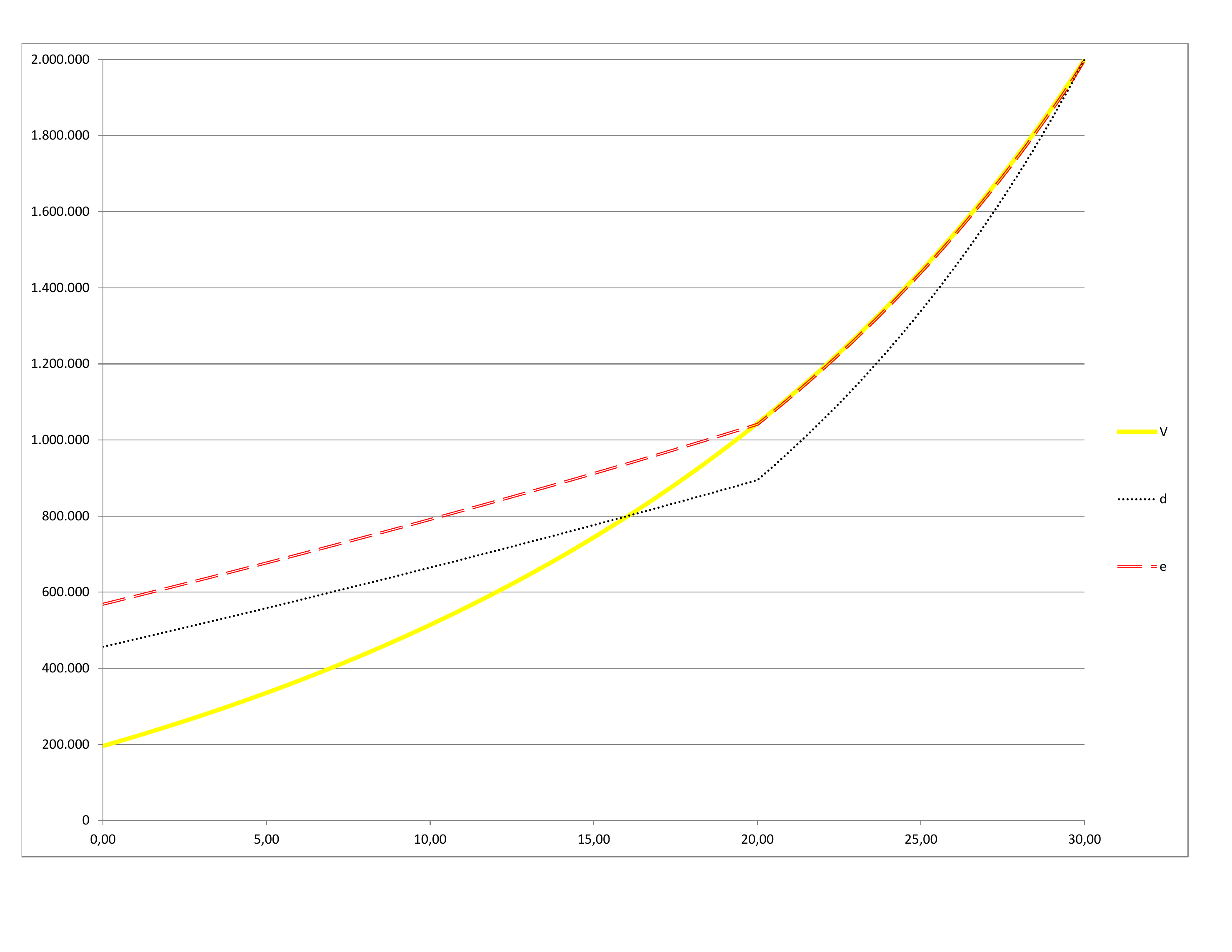}
\caption{Example 4. The technical interest rate is $\hat{r}=0.05$. The
market interest rate is $r(t)=0.01\cdot 1_{(t\leq 20)}+0.065\cdot 1_{(t>20)}$. After time $t=20$ it is optimal to surrender. Before time $t=20$ it is wise
to plan to surrender at time $t=20$.}
\label{fig:model4}      
\end{figure}

\paragraph{Example 4: Market interest rate is increasing}
 Assume $r(t)=0.01\cdot 1_{(t\leq 20)}+0.065\cdot 1_{(t>20)}$. The qualitative feature we capture is that the interest rate crosses the guaranteed interest rate upwards. The reserve
developments are displayed in Figure \ref{fig:model4}. In this situation we
have that after time $t=20$ it is optimal to surrender. Before time $t=20$
it is optimal to plan to surrender at time $t=20$. With this strategy the
policyholder benefits from both the high market interest rate after time $t=20$ and the technical interest rate before time $t=20$ when the market
interest rate is low. Thereby, unlike in the previous three examples, the
worst case reserve is no longer the supremum of the surrender value and the
market reserve with no surrender. Before time $t=20$ the worst case reserve
is higher than both of the other reserves, because there exists a surrender
strategy which is better for the policyholder than both immediate surrender
and no surrender.

We recall that the reserves of Model a and Model b converge to the worst
case reserve when the rationality parameter converges to infinity. Thus, if
the rationality parameter is sufficiently high and the future increase in
interest rate is sufficiently high, then the reserves of Model a and Model b
become higher than the maximum of the surrender value and the market reserve
of Model d with no surrender. 

\appendix
\section{Proof of Theorem \ref{thm:conv}}
The proof is divided in two parts. One part associated with the
risk from the $\nu _{\theta }$ based stopping time surrendering before the
optimal time $u^{\ast }$ and another part associated with the risk from the $%
\nu _{\theta }$ based stopping time surrendering after the optimal time $%
u^{\ast }$. For this reason we define an intermediate reserve, $W_{\theta }$%
. The surrender strategy related to $W_{\theta }$ resembles the one related
to $V_{\theta }$. The only difference is that the strategy related to $%
W_{\theta }$ does not surrender before the optimal time. Mathematically we
make the following definition. Let $\hat{\tau}_{\theta ,t}$ be a stopping
time for which the policyholder surrenders at time $u$ with intensity $\nu
_{\theta }(u)1_{(u\geq u^{\ast }(t))}$. We may write this stopping time in
a convenient way by introducing stopping times, $\hat{\tau}_{\theta ,t}^{i}$%
, given recursively by $\hat{\tau}_{\theta ,t}^{0}\equiv 0$ and $\hat{\tau}%
_{\theta ,t}^{i}$ for $i\in {\mathbb{N}}$ surrenders with intensity $\nu
_{\theta }(u)1_{(u\geq \hat{\tau}_{\theta ,t}^{i-1})}$. With these
definitions we get: 
\begin{equation}\nonumber
(I,\hat{\tau}_{\theta ,t})\overset{d}{=}(I,\sum_{i=1}^{\infty }\hat{\tau}%
_{\theta ,t}^{i}1_{(\hat{\tau}_{\theta ,t}^{i-1}<u^{\ast }(t)\leq \hat{\tau}%
_{\theta ,t}^{i})}).
\end{equation}
This identity comes from renewal theory and the memoryless property of the
exponential distribution. It says that it does not matter if we set the
surrender intensity to zero before the optimal time or if we make the policy
holder regret her decision every time she is about to surrender before the
optimal time. We denote for $s\in \lbrack t,n]$ by $W_{\theta }(t,s)$ the
reserve at time $s$ associated with the surrender strategy $\hat{\tau}%
_{\theta ,t}$. Then, from the identity above we find that%
\begin{equation}\nonumber
W_{\theta }(t,s)=V(s)+\sum_{i=1}^{\infty }{\mathbb{E}}_{s}\left[
e^{-\int_{s}^{\hat{\tau}_{\theta }^{i}}r(u)+\mu (u)du}(G(\hat{\tau}_{\theta
}^{i})-V(\hat{\tau}_{\theta }^{i}))1_{(\hat{\tau}_{\theta }^{i-1}\leq
u^{\ast }(t)<\hat{\tau}_{\theta }^{i})}\right] .
\end{equation}%805-806
\paragraph{Part 1:} 
First we show that for every $t\in \lbrack 0,n]$: 
\begin{equation}\nonumber
\liminf_{\theta \rightarrow \infty }V_{\theta }(t)\leq \liminf_{\theta
\rightarrow \infty }W_{\theta }(t,t).
\end{equation}
To prove this we use, given $t\in \lbrack 0,n]$ and $\varepsilon >0$, the
following notation about stopping times, $\tau $: 
\begin{eqnarray*}
\{\tau \ good\} &=&\{G(\tau )-V_{\theta }(\tau )\geq 0\}, \\
\{\tau \ ok\} &=&\{G(\tau )-V_{\theta }(\tau )\in \lbrack -\varepsilon ,0)\},
\\
\{\tau \ bad\} &=&\{G(\tau )-V_{\theta }(\tau )<-\varepsilon \}.
\end{eqnarray*}
Thus, a stopping time, $\tau $, is called good when it is profitable to
surrender at the corresponding time, and it is called bad when the policy
holder loses more than $\varepsilon $ on surrendering. In the following, let 
$u^{\ast }\equiv u^{\ast }(t)$ and let $\hat{\tau}_{\theta }^{i}\equiv \hat{%
\tau}_{\theta ,t}^{i}$. By induction one can show that for every $m\in {%
\mathbb{N}}$: 
\begin{eqnarray}
V_{\theta }(t) &=&V(t)+{\mathbb{E}}_{t}\left[ e^{-\int_{t}^{\hat{\tau}%
_{\theta }^{1}}r(u)+\mu (u)du}(G(\hat{\tau}_{\theta }^{1})-V(\hat{\tau}%
_{\theta }^{1}))\right]   \nonumber \\
&\geq &V(t)\nonumber\\
&&+\sum_{i=1}^{m}{\mathbb{E}}_{t}\left[ e^{-\int_{t}^{\hat{\tau}%
_{\theta }^{i}}r(u)+\mu (u)du}(G(\hat{\tau}_{\theta }^{i})-V(\hat{\tau}%
_{\theta }^{i}))1_{(\hat{\tau}_{\theta }^{i-1}\leq u^{\ast }<\hat{\tau}%
_{\theta }^{i},\hat{\tau}_{\theta }^{1},\ldots ,\hat{\tau}_{\theta
}^{i-1}ok\ or\ good)}\right]   \nonumber \\
&&+{\mathbb{E}}_{t}\left[ e^{-\int_{t}^{\hat{\tau}_{\theta }^{m+1}}r(u)+\mu
(u)du}(G(\hat{\tau}_{\theta }^{m+1})-V(\hat{\tau}_{\theta }^{m+1}))1_{(\hat{%
\tau}_{\theta }^{m}\leq u^{\ast },\hat{\tau}_{\theta }^{1},\ldots ,\hat{\tau}%
_{\theta }^{m}ok\ or\ good)}\right]   \nonumber \\
&&+\sum_{i=1}^{m}{\mathbb{E}}_{t}\left[ e^{-\int_{t}^{\hat{\tau}_{\theta
}^{i}}r(u)+\mu (u)du}(G(\hat{\tau}_{\theta }^{i})-V(\hat{\tau}_{\theta
}^{i}))1_{(\hat{\tau}_{\theta }^{i}\leq u^{\ast },\hat{\tau}_{\theta
}^{1},\ldots ,\hat{\tau}_{\theta }^{i-1}ok\ or\ good,\hat{\tau}_{\theta
}^{i}bad)}\right]   \nonumber \\
&&-\varepsilon \sum_{i=1}^{m}{\mathbb{E}}_{t}\left[ e^{-\int_{t}^{\hat{\tau}%
_{\theta }^{i}}r(u)+\mu (u)du}1_{(\hat{\tau}_{\theta }^{i}\leq u^{\ast },%
\hat{\tau}_{\theta }^{1},\ldots ,\hat{\tau}_{\theta }^{i-1}ok\ or\ good,\hat{%
\tau}_{\theta }^{i}ok)}\right] \label{eq:induc1}
\end{eqnarray}
The idea is that the reserve $V_{\theta }$ corresponds to the technical
reserve, $V$, plus the expected gain from surrender. We investigate what
happens if the policyholder regrets to surrender. The impact if the policy
holder regrets to surrender at the observed stopping time, $\hat{\tau}%
_{\theta }^{1}$, depends on whether this stopping time was good, ok or bad.
If the stopping time is good, then we know that the value of the gain of
surrender is at least as high as waiting for the next time to surrender, and
if the stopping time is ok, then we know that the value of the gain of
surrender is at most $\varepsilon $ worse than waiting for the next time to
surrender. In the above expression we have made these judgements for up to $m
$ surrender possibilities before the optimal time.

The sum in the first line corresponds to the case when one of the first $m$
stopping times reaches beyond the optimal time, $u^{\ast }$. The terms of
the second line correspond to the case when all of the first $m$ stopping
times are before the optimal time, $u^{\ast }$, and they have all been ok or
good. In this case, the value of the gain of surrendering at the first
stopping time is no higher than waiting for the $m+1$'th stopping time. The
sum in the third line corresponds to the case when one of the $m$ first
stopping times is bad and is before the optimal time, $u^{\ast }$. The sum
of the fourth line is a correction of the $\varepsilon $-small loses from ok
stopping times.

If we display the bound relative to $W_{\theta }$ instead of relative to the
technical reserve, $V$, then we get the following expression: 
\begin{eqnarray*}
&&V_{\theta }(t) \\
&\geq &W_{\theta }(t,t)-\sum_{i=1}^{\infty }{\mathbb{E}}_{t}
\left[ e^{-\int_{t}^{\hat{\tau}_{\theta }^{i}}r(u)+\mu (u)du}(G(\hat{\tau}
_{\theta }^{i})-V(\hat{\tau}_{\theta }^{i}))1_{(\hat{\tau}_{\theta
}^{i-1}\leq u^{\ast }<\hat{\tau}_{\theta }^{i},\exists j\in \{1,\ldots
,i-1\}:\ \hat{\tau}_{\theta }^{j}bad)}\right]  \\
&&-\sum_{i=m+1}^{\infty }{\mathbb{E}}_{t}\left[ e^{-\int_{t}^{\hat{\tau}
_{\theta }^{i}}r(u)+\mu (u)du}(G(\hat{\tau}_{\theta }^{i})-V(\hat{\tau}
_{\theta }^{i}))1_{(\hat{\tau}_{\theta }^{i-1}\leq u^{\ast }<\hat{\tau}
_{\theta }^{i},\hat{\tau}_{\theta }^{1},\ldots ,\hat{\tau}_{\theta
}^{i-1}ok\ or\ good)}\right]  \\
&&+{\mathbb{E}}_{t}\left[ e^{-\int_{t}^{\hat{\tau}_{\theta }^{m+1}}r(u)+\mu
(u)du}(G(\hat{\tau}_{\theta }^{m+1})-V(\hat{\tau}_{\theta }^{m+1}))1_{(\hat{
\tau}_{\theta }^{m}\leq u^{\ast },\hat{\tau}_{\theta }^{1},\ldots ,\hat{\tau}
_{\theta }^{m}ok\ or\ good)}\right]  \\
&&+\sum_{i=1}^{m}{\mathbb{E}}_{t}\left[ e^{-\int_{t}^{\hat{\tau}_{\theta
}^{i}}r(u)+\mu (u)du}(G(\hat{\tau}_{\theta }^{i})-V(\hat{\tau}_{\theta
}^{i}))1_{(\hat{\tau}_{\theta }^{i}\leq u^{\ast },\hat{\tau}_{\theta
}^{1},\ldots ,\hat{\tau}_{\theta }^{i-1}ok\ or\ good,\hat{\tau}_{\theta
}^{i}bad)}\right]  \\
&&-\varepsilon \sum_{i=1}^{m}{\mathbb{E}}_{t}\left[ e^{-\int_{t}^{\hat{\tau}
_{\theta }^{i}}r(u)+\mu (u)du}1_{(\hat{\tau}_{\theta }^{i}\leq u^{\ast },
\hat{\tau}_{\theta }^{1},\ldots ,\hat{\tau}_{\theta }^{i-1}ok\ or\ good,\hat{
\tau}_{\theta }^{i}ok)}\right] .
\end{eqnarray*}
In the limit of $\theta $, $\varepsilon $ and $m$, then $W_{\theta }$ is the
only term which does not converge to 0. To see this, notice that there
exists some $K>0$ such that for all $u\in \lbrack t,n]$: 
\begin{equation}\nonumber
G(u)-V(u),\in \lbrack -K,K],\ \ \text{ and }\ \ G(u)-V_{\theta }(u)\in
\lbrack -K,K].
\end{equation}
That is, for any stopping time, the adjustment $G-V$ is bounded by $K$.
Thereby we may further bound the value of $V_{\theta }$ by replacing each of
these adjustments with $-K$ times an upper bound of the probability of the
corresponding event: 
\begin{eqnarray*}
V_{\theta }(t) &\geq &W_{\theta }(t,t)-K{\mathbb{P}}_{t}(\exists j\in {%
\mathbb{N}}:\ \hat{\tau}_{\theta }^{j}bad,\hat{\tau}_{\theta }^{j}\leq
u^{\ast })-K\sum_{i=m+1}^{\infty }{\mathbb{P}}_{t}(\hat{\tau}_{\theta
}^{i-1}\leq u^{\ast }<\hat{\tau}_{\theta }^{i}) \\
&&-K{\mathbb{P}}_{t}(\hat{\tau}_{\theta }^{m}\leq u^{\ast },\hat{\tau}%
_{\theta }^{1},\ldots ,\hat{\tau}_{\theta }^{m}ok\ or\ good)-K{\mathbb{P}}%
_{t}(\exists j\in {\mathbb{N}}:\ \hat{\tau}_{\theta }^{j}bad,\hat{\tau}%
_{\theta }^{j}\leq u^{\ast }) \\
&&-\varepsilon \sum_{i=1}^{m}{\mathbb{P}}_{t}(\hat{\tau}_{\theta }^{i}\leq
u^{\ast },\hat{\tau}_{\theta }^{i}ok) \\
&\geq &W_{\theta }(t,t)-K(1-e^{(n-t)\bar{h}_{\theta }(-\varepsilon
)})-K\sum_{i=m+1}^{\infty }{\mathbb{P}}_{t}(\hat{\tau}_{\theta }^{i-1}\leq
u^{\ast }<\hat{\tau}_{\theta }^{i}) \\
&&-K{\mathbb{P}}_{t}(\hat{\tau}_{\theta }^{m}\leq u^{\ast })-K(1-e^{(n-t)%
\bar{h}_{\theta }(-\varepsilon )})-\varepsilon n. \\
&&
\end{eqnarray*}
Given $\theta $ and $\varepsilon $, then this holds for every $n$. Thus, the
second sum can be made arbitrarily small and so can ${\mathbb{P}}_{t}(\hat{%
\tau}_{\theta }^{n}\leq u^{\ast })$, the later follows because given $\theta 
$, then the intensity of surrender is bounded on $[0,n]$ and thus the
distribution of the number of $\hat{\tau}_{\theta }^{i}$ before $u^{\ast }$
is bounded by a Poisson distribution. Thereby: 
\begin{equation}\nonumber
\liminf_{\theta \rightarrow \infty }V_{\theta }(t)\geq \liminf_{\theta
\rightarrow \infty }W_{\theta }(t,t).
\end{equation}
We find from the calculations above that the lower bound holds because the
surrender strategy of $W_{\theta }$ and $V_{\theta }$ only differs by the
strategy of $W_{\theta }$, regretting every surrender before the optimal
time. The impact of this difference is bounded because the following main
reasons: The probability of a bad stopping time converges to zero in the
limit because of (\ref{eq:conv1}). The number of ok or good stopping times
occurring before the optimal time is finite. Regret of a good stopping time
decreases the value. Regret of an ok stopping time has an impact bounded by $%
\varepsilon $. At last, the technical calculations justify that the
convergence of $\varepsilon $ does not cancel the impact of the convergence
of (\ref{eq:conv1}).

\paragraph{Part 2:} 
Consider some arbitrary $t\in \lbrack 0,n]$. We wish to show that: 
\begin{equation}\nonumber
W_{\theta }(t,t)\rightarrow W(t),\ \ \ \theta \rightarrow.
\end{equation}
Let $u^{\ast }\equiv u^{\ast }(t)$ and $\hat{\tau}_{\theta }=\hat{\tau}%
_{\theta ,t}$, and notice that since the policyholders related to $W_{\theta
}$, $W$ and $V_{\theta }$ behave similarly before time $u^{\ast }$, then
convergence at time $t$ corresponds to convergence at time $u^{\ast }$. This
is seen from: 
\begin{eqnarray*}
W(t)-W_{\theta }(t,t) &=&{\mathbb{E}}_{t}\left[ e^{-\int_{t}^{u^{\ast
}}r(u)+\mu (u)du}(G(u^{\ast })-V(u^{\ast }))\right.  \\
&&\left. -e^{-\int_{t}^{\hat{\tau}_{\theta }}r(u)+\mu (u)du}(G(\hat{\tau}
_{\theta })-V(\hat{\tau}_{\theta }))\right]  \\
&=&e^{-\int_{t}^{u^{\ast }}r(u)+\mu (u)du}(W(u^{\ast })-W_{\theta }(u^{\ast
},u^{\ast })) \\
&=&e^{-\int_{t}^{u^{\ast }}r(u)+\mu (u)du}(W(u^{\ast })-V_{\theta }(u^{\ast
})).
\end{eqnarray*}
Thereby, it is sufficient to prove that $V_{\theta }(u^{\ast })\rightarrow
W(u^{\ast })$ when $\theta \rightarrow \infty $. Either this holds, or there
is some $\varepsilon _{1}>0$ and some sequence $(\theta _{i})_{i\in {\mathbb{
N}}}$ converging to infinity such that for all $i\in {\mathbb{N}}$: $
V_{\theta _{i}}(u^{\ast })<W(u^{\ast })-2\varepsilon _{1}$. Thereby $
V_{\theta _{i}}(u^{\ast })<G(u^{\ast })-2\varepsilon _{1}$.

The derivative of $V_{\theta_i}$ is uniformly bounded over $i$ as long as $
V_{\theta_i}<G$. Thus, there exists some $\delta_{1}$ such that $
V_{\theta_i}(u)\leq G(u)-\varepsilon_1$ for $u\in[u^*,u^*+\delta_1]$. For
this time interval the gain of surrender compared to waiting is at least $
\varepsilon_1$, and thereby, for this time interval the intensity for
surrender is at least $\underline{h}_{\theta_i}(\varepsilon_1)$. 

As $V$ is continuous, then, for every $\varepsilon_2>0$, there exists some $
\delta_{2}$ such that $(G(u^*)-V(u^*))-e^{-\int_{u^*}^{t}r(u)+
\mu(u)du}(G(t)-V(t))\leq \varepsilon_2$ for $t\in[u^*,u^*+\delta_{2}]$. That
is, if surrender happens within time $\delta_2$ of the optimal time then the
loss of the delay is at most $\varepsilon_2$.

Now, let $\delta =\delta _{1}\wedge \delta _{2}$. Then the loss of surrender
according to $\hat{\tau}_{\theta _{i}}$ instead of at the optimal time is
bounded in the following way: 
\begin{eqnarray*}
&&W(u^{\ast })-V_{\theta _{i}}(u^{\ast }) \\
&=&{\mathbb{E}}\left[ \left. (G(u^{\ast })-V(u^{\ast }))-(G(\hat{\tau}
_{\theta _{i}})-V(\hat{\tau}_{\theta _{i}}))e^{-\int_{u^{\ast }}^{\hat{\tau}
_{\theta _{i}}}r(x)+\mu (x)dx}\right\vert \hat{\tau}_{\theta _{i}}\leq
\delta \right] {\mathbb{P}}(\hat{\tau}_{\theta _{i}}\leq \delta ) \\
&&+{\mathbb{E}}\left[ \left. (G(u^{\ast })-V(u^{\ast }))-(G(\hat{\tau}
_{\theta _{i}})-V(\hat{\tau}_{\theta _{i}}))e^{-\int_{u^{\ast }}^{\hat{\tau}
_{\theta _{i}}}r(x)+\mu (x)dx}\right\vert (\hat{\tau}_{\theta _{i}}\leq
\delta )^{c}\right] {\mathbb{P}}((\hat{\tau}_{\theta _{i}}\leq \delta
)^{c}) \\
&\leq &\varepsilon _{2}+{\mathbb{E}}\left[ \left. (G(u^{\ast })-V(u^{\ast
}))-(G(\hat{\tau}_{\theta _{i}})-V(\hat{\tau}_{\theta
_{i}}))e^{-\int_{u^{\ast }}^{\hat{\tau}_{\theta _{i}}}r(x)+\mu
(x)dx}\right\vert (\hat{\tau}_{\theta _{i}}\leq \delta )^{c}\right]
e^{-\delta \underline{h}_{\theta _{i}}(\varepsilon _{1})} \\
&\leq &\varepsilon _{2}+2Ke^{-\delta \underline{h}_{\theta
_{i}}(\varepsilon _{1})}.
\end{eqnarray*}
Thus $V_{\theta _{i}}(u^{\ast })\rightarrow W(u^{\ast })$ as $\theta
\rightarrow \infty $, and the result follows.


\begin{thebibliography}{}

\bibitem{Ba03}
Bacinello AR (2003) 
Fair Valuation of a Guaranteed Life Insurance Participating Contract Embedding a Surrender Option.
J Risk Insur 70(3):461--487. doi: 10.1111/1539-6975.t01-1-00060

\bibitem{BM13}
Buchardt K, M\o ller T (2013) 
Life insurance cash flows with policyholder
behaviour, 
Preprint (submitted), available at
http://www.math.ku.dk/ buchardt/

\bibitem{BMS13}
Buchardt K, M\o ller T, Schmidt KB (2013) 
Cash flows and policyholder behaviour in the semi-Markov life insurance setup. Scand Actuar J 6:765--798

\bibitem{EK14}
Eling M,  Kiesenbauer D (2014) 
What Policy Features Determine Life Insurance Lapse? An analysis of the German Market.
J Risk Insur 81(2):241--269. doi: 10.1111/j.1539-6975.2012.01504.x

\bibitem{FV02}
Forsyth PA, Vetzal KR (2002) 
Quadratic convergence for valuing American options using a penalty method.
SIAM J Sci Comput 23(6):2095--2122. doi: 10.1137/S1064827500382324

\bibitem{GP14}
Gad KST, Pedersen JL (2014). 
Profit dependent Exercise of the American Put,
Preprint, available at arxiv.org/abs/1410.1287 

\bibitem{Gi10}
De Giovanni D (2010). 
Lapse rate modeling: a rational expectation approach. Scand Actuar J 2010(1):56--67

\bibitem{GJ00}
Grosen A, J\o rgensen PL (2000). 
Fair Valuation of Life Insurance
Liabilities: The Impact of Interest Rate Guarantees, Surrender Options, and
Bonus Policies. Insur Math Econ 26(1):37--57

\bibitem{HNSS14} 
Henriksen LFB, Nielsen JW, Steffensen M, Svensson C (2014). 
Markov chain modeling of policyholder behaviour in life insurance and pension.
Eur Actuar J 4(1):1--29. doi: 10.1007/s13385-014-0091-2

%bog
\bibitem{KYPR06}
Kyprianou A. E. (2006)
Introductory Lectures on Fluctuations of Levy Processes with Applications.
Springer %,place

%bog
\bibitem{TMMS07}
M\o ller T, Steffensen M (2007) 
Market-Valuation Methods in Life and Pension Insurance.
Cambridge %, place

\bibitem{ANS78}
Shiryayev AN (1978) 
Optimal Stopping Rules.
Springer-Verlag%, place

\bibitem{MS02} 
Steffensen M. (2002) 
Intervention options in life insurance.
Insur Math Econ 31:71--85
\end{thebibliography}
\end{document}